\documentclass[aps, prb , 10pt ,a4paper,twocolumn,floatfix,showpacs,showkeys,amsmath,amssymb,altaffilletter]{revtex4-1}
\usepackage{graphicx}

\usepackage{epstopdf,amsmath,amssymb, booktabs}

\newcommand{\beq}{\begin{equation}}
\newcommand{\eeq}{\end{equation}}
\newcommand{\bea}{\begin{eqnarray}}
\newcommand{\eea}{\end{eqnarray}}
\renewcommand{\d}[1]{\ensuremath{\,\rm d#1}}

\newcommand{\chem}[1]{\ensuremath {\rm{#1}}}
\newcommand{\plq}[0]{PL\textsuperscript{Q}}
\newcommand{\plnq}[0]{PL\textsuperscript{nQ}}

\begin{document}

\title{Investigation of exciton transport in crystalline thin-films of the organic semiconductor di-indeno-perylene using photoluminescence analyses}%

\author{A. K. Topczak,$^{1}$ T. Roller,$^{2}$ B. Engels,$^{1}$ W. Br\"utting,$^{3}$ J. Pflaum$^{1,4}$}
\email{jpflaum@physik.uni-wuerzburg.de}
\affiliation{$^{1}$Julius-Maximillians University, Am Hubland, 97074 W\"urzburg  Germany\linebreak
$^{2}$University of Stuttgart, Pfaffenwaldring 57, 70569 Stuttgart, Germany\linebreak
$^{3}$University of Augsburg, Universit\"atsstraße 1, 86159 Augsburg, Germany\linebreak
$^{4}$ZAE Bayern, Am Hubland, 97074 W\"urzburg  Germany }
\date{\today}%

\begin{abstract}
The exciton transport in the prototypical organic semiconductor Di-Indeno-Perylene (DIP) has been investigated by means of photoluminescence (PL) quenching and interpreted by an advanced exciton diffusion model including interference effects, quencher penetration as well as non-perfect exciton quenching. X-ray diffraction revealed a correlation between the exciton diffusion length of about 90 nm and the structural coherence length of the DIP layers. Temperature dependent studies in a range of 5 - 300 K indicated an incoherent exciton transport above 80 K at activation energies of 10 - 20 meV related to the thickness dependent gradient of exciton density. Below 80 K a coherent exciton transport can be observed by the reduced phonon-interaction at cryogenic temperatures. This manuscript is a pre-final version.
\end{abstract}
\maketitle

By their progressively improving power conversion efficiency, organic thin film photovoltaic cells pave their way towards first applications \cite{greggpaper}. In organic bilayer photovoltaic cells, the competition between exciton diffusion and absorption length is decisive for the applicable thickness of the photo-active layers and thereby the overall device efficiency \cite{peumansyakimov}. As it became obvious from observations of band-like, coherent exciton motion in organic single crystals \cite{orritbernard} that morphology significantly influences the diffusion characteristics of optical excitations, molecular model systems supporting coherent exciton transport by their crystalline film structure are highly desired \cite{{luntbenzinger},{kurrlepflaum}}. To aim for this challenge we have performed temperature dependent exciton transport studies in thin films of the prototypical organic semiconductor Di-Indeno-Perylene (DIP).

Bilayer thin films cells of DIP in combination with the electron acceptor \chem{C_{60}} have shown superior performance including fill factors above 70 \% and power conversion efficiencies of up to 4 \% \cite{wagnergruber}. 
 
\begin{figure} [h]
\centering
\includegraphics[width=0.90\columnwidth]{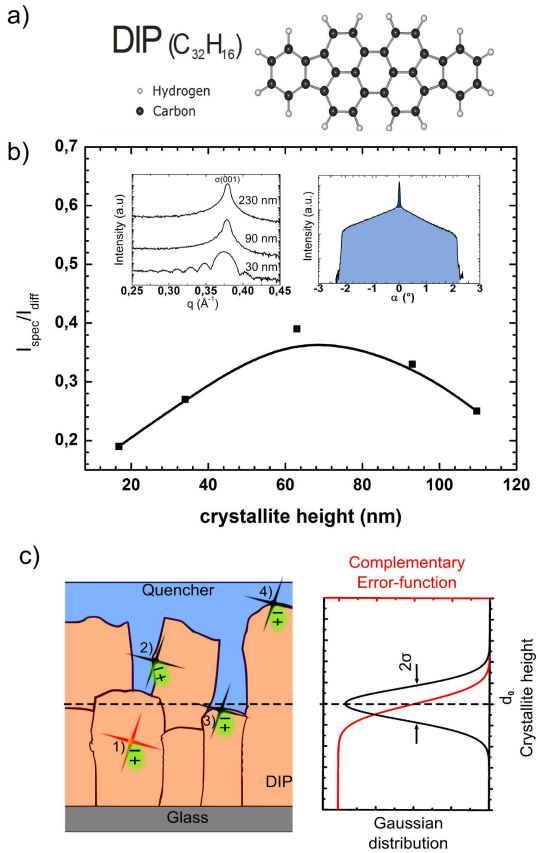}
\caption{(a) Structural model of DIP. (b) X-ray structural analyses and Laue-oscillations (upper left inset) indicating the high crystallinity of the DIP films. The ratio between specular intensity (dark blue area) and the diffuse scattering intenity (light blue area) has been considered as a measure for the structural quality of the DIP layers (upper right inset). (c) Model of the DIP film texture together with possible exciton recombination processes. Integration of the Gaussian distribution of the DIP crystallite heights (black curve) yields the complementary error function (red curve).}
\label{fig:fig1}
\end{figure}

Therefore, in this paper we present an approach, including morphological aspects as boundary effects, quencher penetration and optical interference effects, to determine the excitonic transport properties by thickness dependent Photoluminescence (PL) studies. This method also offers insight in the dominate excitonic transport mechanisms by temperature dependent PL-studies. For this purpose, DIP samples were prepared by organic molecular beam epitaxy on cleaned glass slides. The resulting film structure of the samples was analyzed in detail by X-ray diffraction at various geometries. Bragg-Brentano scans confirmed the previously reported $\left(001\right)$ lattice spacing of about 1.66 nm corresponding to an almost upright orientation of the molecules \cite{duerrschreiber} and the long-range coherence of the lattice planes is indicated by pronounced Laue oscillations at the $\left(100\right)$ Bragg peak (left inset of fig. \ref{fig:fig1}b). Furthermore, DIP film mosaicity is charaterized by an averaged crystallite tilting of only 0.04 deg along the surface normal, i.e. along the exciton transport direction (right inset of fig. \ref{fig:fig1}b). To define a measure for the film quality we employed the ratio of specular versus diffuse intensity (dark vs. light blue shaded areas in the right inset of fig. \ref{fig:fig1}b), the latter being influenced by structural inhomogeneities within the organic film \cite{nickelbarabash}. As can be seen, this ratio peaks at a thickness of about 70 nm in agreement with the crystallite height, determined by Laue-oscillations. Based on this data a model of the underlying film structure has been deduced and is described by a complementary error function representing the integrated Gaussian distribution of DIP crystallite heights (fig. \ref{fig:fig1}c).

\begin{figure} [h]
\centering
\includegraphics[width=0.90\columnwidth]{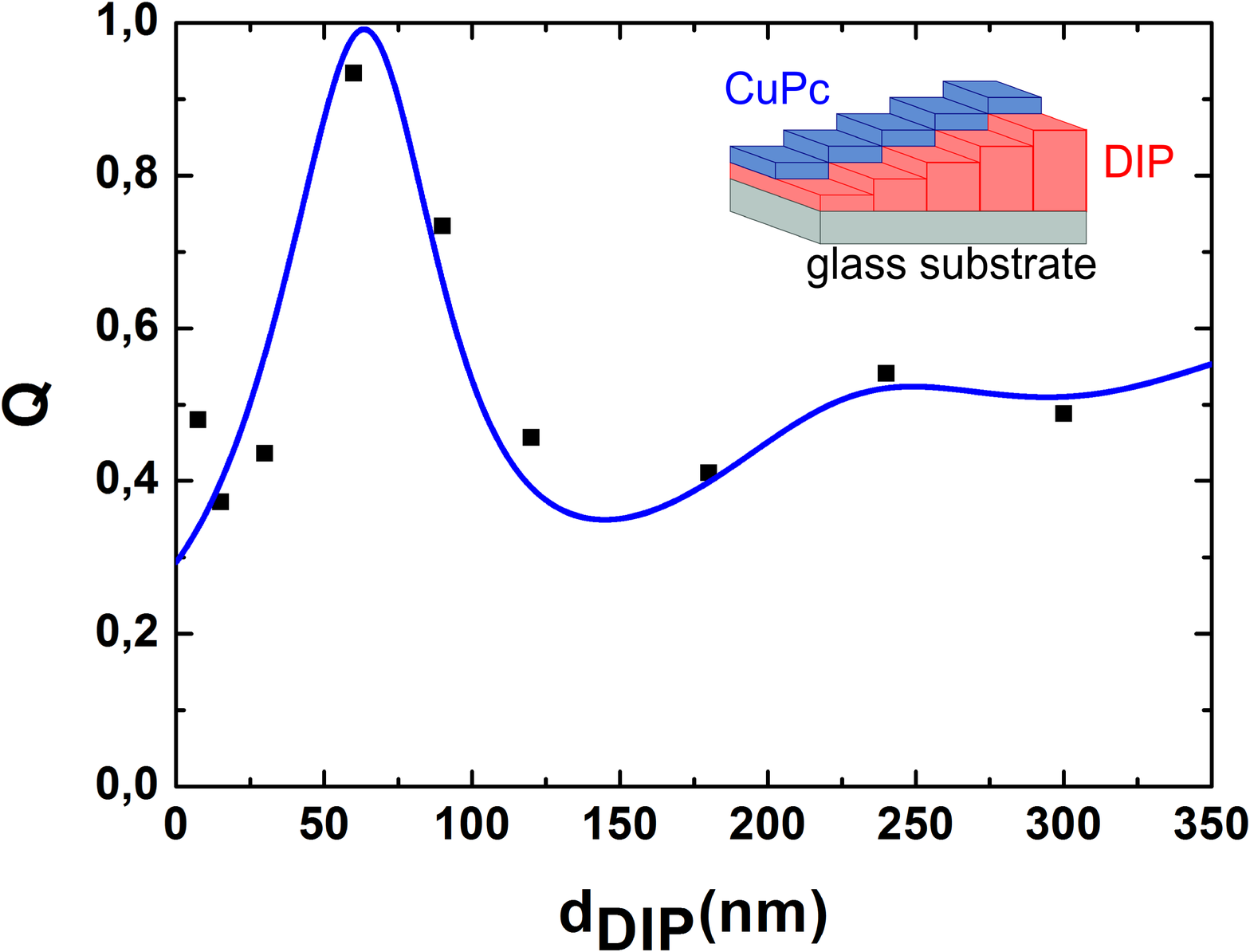}
\caption{Thickness dependent quenching ratio Q=$PL^Q/PL^{nQ}$. The fit (blue line) corresponds to the model discussed in text accounting for interference effects and interdiffusion at the DIP/quencher interface. The inset displays the sample structure.}
\label{fig:fig2}
\end{figure}

For the Photoluminescence (PL) studies DIP thin films were prepared with Copper Phthalocyanine (CuPc) thermally evaporated on one half of the DIP stepped wedge (inset of fig. \ref{fig:fig2}), acting as exciton quencher due to its HOMO and LUMO positions of 5.1 eV and 3.2 eV with respect to those of DIP, 5.4 eV and 2.9 eV \cite{{kahnkoch}, {wagnergruber}}. The PL intensity of the DIP films, illuminated through the substrate by a NdYag-Laser at 532nm, was measured by a sensitive CCD detector after passing a 590 nm long pass filter to block background intensity of the excitation. For a given DIP layer thickness, the comparison of the relative PL intensity ratio Q=$PL^Q/PL^{nQ}$, recorded on the bare ($PL^{nQ}$) and the CuPc-covered ($PL^Q$) area reveals contributions from non-radiative decay channels and dissociation of excitons \cite{peumansyakimov}. In films thinner than $L_D$ most of the excitons will reach the quenching interface within their radiative lifetime, thus strongly reducing the PL signal. On the other hand for films thicker than $L_D$ the PL signal is expected to dependent only on the absorption and therefore should be identical for the free and the quencher-covered areas. As a result, $L_D$ in the molecular layer can be determined by thickness dependent PL measurements. It is important to stress at this point that according to the layer thicknesses and $L_D$ under study  \cite{scullymcgehee} as well as by the small spectral overlap for DIP, contributions to quenching by Fluorescence Resonance Energy Transfer (FRET) can be neglected \cite{luntgiebink}.

However, the situation becomes more complex for real samples, especially if the absorption layer thickness approaches the excitation wavelength within the organic material. For thicknesses equal or larger than $n\lambda/4$ interference effects have to be taken into account \cite{theanderyartsev}, requiring detailed information on the indicatrix of the organic material. Therefore, to quantify exciton quenching one has to calculate at first the intensity profile of incident light across the sample yielding the initial exciton density. Accounting for light reflection at the organic interfaces yields the depth dependent intensity profile \cite{{theanderyartsev},{stubingerbrutting}}

\bea
{I(z)\over I_0}  &=&   e^{-\alpha z} + \rho^2 e^{-\alpha (2d-z)}   \nonumber \\
								 &+& 2 \rho e^{-\alpha d}\cos (-2k(d-z)-\delta) \label{eq::I}
\eea

Here z describes the position in the film (z = 0 refers to the glass/organic-interface), $\alpha$ is the absorption coefficient, $\rho$ is the reflection coefficient at the organic interface, d is the sample thickness,  k is the wave vector and $\delta$ is the phase shift occurring upon reflection at the backside of the organic layer. With {\it n(z,t)} being the local exciton density at time t, the continuity equation reads

\beq
\frac{\d n(z,t)}{\d t} = D\frac{\partial^2n(z,t)}{\partial z^2}-\frac{n(z,t)}{\tau}+G(z,t) \label{eq::n}
\eeq

For given exciton lifetime $\tau$, the exciton generation term {\it G(z,t)} under constant illumination, i.e. at steady state conditions ($ \d n(z,t)/\d t = 0$), is expected to be proportional to the light intensity profile. Appropriate boundary conditions can be achieved by neglecting quenching at the glass/organic-interface, $\left.\frac{\partial n(z,t)}{\partial z}\right|_{z=0}=0$ and assuming either complete quenching, $n(z,t) = 0$, or no quenching, $\left.\frac{\partial n(z,t)}{\partial z}\right|_{z=d}=0$, at the organic/quencher- or organic/air-interface, respectively. Real samples will always operate between these two extremes, i.e. partial quenching might occur at the organic/air-interface e.g. due to surface states as well as non-ideal quenching at the organic/quencher-interface. For perfect quenching the absolute value of the slope and accordingly the exciton concentration gradient will be at maximum $S_{max}$ at the quencher interface. This provides a measure for the quenching quality $V = S/ S_{max}$, where S is the slope at the organic/quencher-interface assuming incomplete quenching. Obviously, $V$ assumes values between 0 and 1. 
As for more complex intensity profiles eq.(\ref{eq::n}) can no longer be solved analytically, we employed an ansatz based on superposition of Dirac delta functions $G(z)= \delta(z-z_0)/\tau$ as generation term. The low exciton densities in our PL studies render the diffusion to be concentration independent and therefore $n_\delta(z,z0)$ can be used in the generation profile. A convolution of the light intensity profile, including interference effects and $n_\delta(z,z0)$ results in

\beq
n(z) = \int\limits_0^d{\Phi \cdot I(z_0)\cdot n_\delta(z,z_0)\d z_0}
\eeq

The factor $\Phi$ describes the quantum efficiency of exciton generation upon photon absorption. Assuming non-radiative decay processes in the absorption layer to be of minor importance, the detected PL intensity corresponds to the number of excitons not quenched within their radiative lifetime

\bea
 PL&=& k\cdot \int\limits_0^d I(z_0)\left(1-V\frac{cosh(z_0/L_D)}{cosh(d/L_D)}\right) \d z_0 \label{eq::PL}
\eea

Here we used the common definition of the exciton diffusion length $L_D=\sqrt{D\tau}$. For $V=0$ each exciton contributes to photoluminescence thereby yielding the integrated intensity profile. Using eq.(\ref{eq::PL}) the PL signal can be calculated for any given $I(z_0)$.
The thickness dependent Q measured at room temperature (RT) is displayed in fig. \ref{fig:fig2}. Three distinct features can be clearly identified: First, Q does not approach zero even for $d_{DIP}\rightarrow0$, suggesting imperfect quenching at the DIP/CuPc interface. Second, the oscillatory thickness behavior of Q refers to the presence of interference effects. Third, even for film thicknesses of 300 nm Q is still below 0.6, indicating the existence of a significant, almost thickness independent quenching contribution.

\begin{figure} [h]
\centering
\includegraphics[width=0.90\columnwidth]{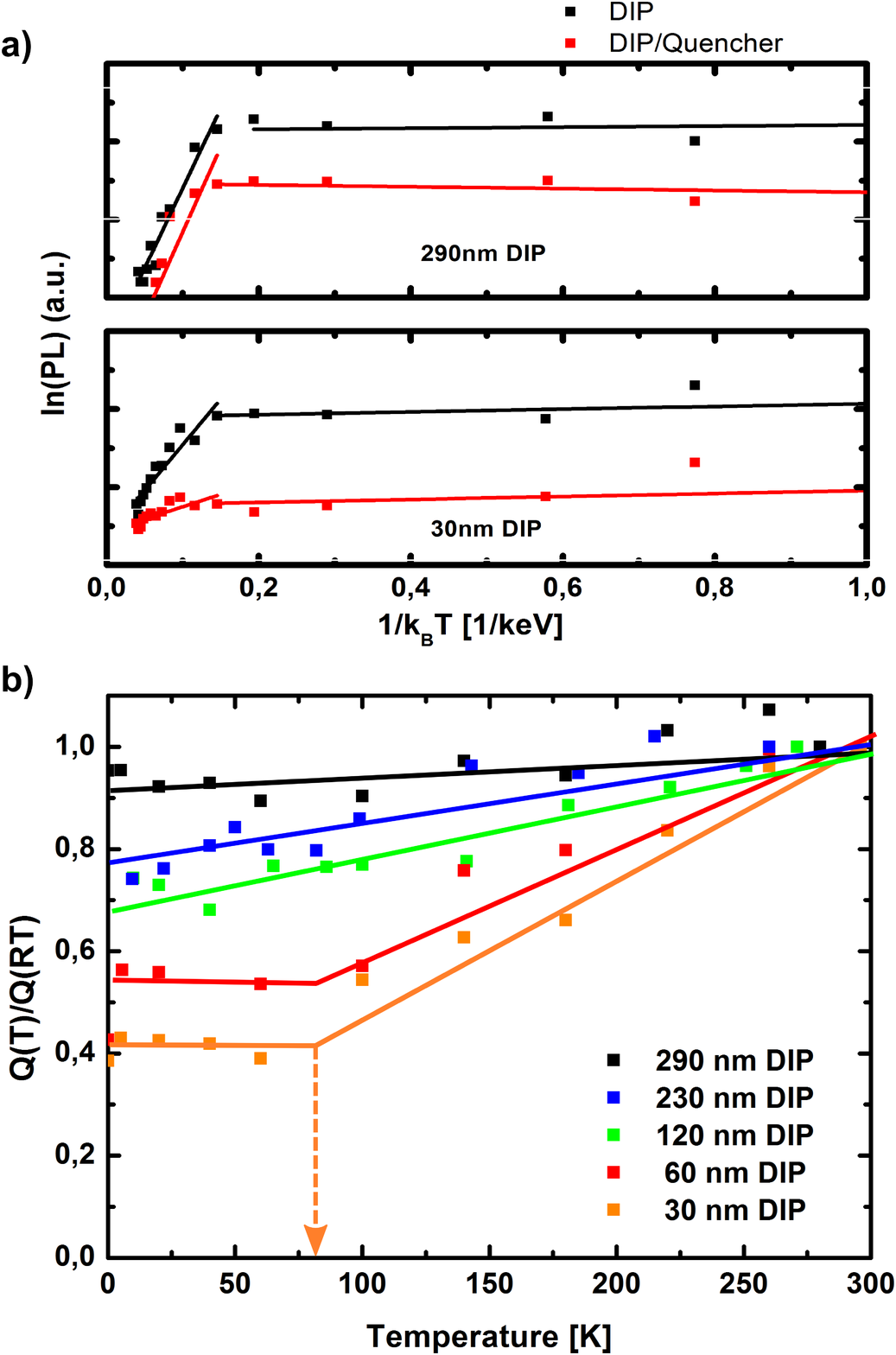}
\caption{(a) Comparison of the PL(T) behavior with and without quenching layer for the thickest and thinnest film of this study. (b) Temperature dependent Q for five representative DIP layer thicknesses normalized to the room temperature value. Thinner layer show the signature of a thermally activated Q above 80 K whereas below this temperature, the ratio remains constant.}
\label{fig:fig3}
\end{figure}

To relate these results to the underlying film morphology, complementary studies on the temperature dependent PL have been carried out on DIP layers of various thicknesses. Fig. \ref{fig:fig3}a shows the PL intensity as a function of the inverse thermal energy recorded on thin (30 nm) and thick (290 nm) DIP layers with an without quenching layer. Below 0.145 $(keV)^{-1}$, i.e. above 80 K, the fluorescence of the free ($\plnq$) and of the CuPc covered ($\plq$) DIP volume is thermally activated, with $\plnq$ showing a distinct increase with thickness from 12 meV to about 21 meV. This tendency can be related to different exciton density gradients as a function of layer thickness. A steeper concentration gradient, occurring for
thinner films, reduces the effective energy required for transport \footnote{Here we refer to an effective activation energy comprising contributions by the concentration gradient as well as by the temperature dependent diffusion constant}. In addition, this is confirmed by the quencher covered 30 nm thick DIP film whose $\plq$ activation energy is about four times smaller than that of the uncovered film. Finally, the value of 21 meV for the thickest DIP film is in good agreement with the PL activation energy of covered DIP irrespective of layer thickness and, therefore, refers to exciton transport within the DIP crystallites. In analogy to charge carrier transport studies on long-range ordered molecular stacks \cite{hannewaldstoja} we attribute the temperature dependent fluorescence behavior to exciton-phonon-interaction constituting a non-radiative decay channel which is suppressed towards lower temperatures, i.e. upon freezing-out non-local phonon modes \cite{matsuipaper}. This assumption is corroborated by the observed activation energy range which coincides with that of non-local phonon modes in polyaromatic single crystals \cite{{ortmannbechstedt},{hannewaldstoja}}. Furthermore, data for all thicknesses in fig. \ref{fig:fig3} reveals the PL intensity to become temperature independent below 80 K. Caused by an almost coherent transport of the (free) excitonic species within the crystalline DIP fraction \cite{munnsiebrand} without dissipative effects by static inhomogenities. 

As it becomes obvious from fig. \ref{fig:fig3}b, the thickness dependent Q is characterized by a temperature dependent decrease for thin DIP films followed by an temperature independent regime below 80K. In contrast, thick films ($d>100nm$) show an almost constant Q over the entire temperature range. This points directly at the morphological origin of this phenomenon which can be interpreted by the previously discussed spatial extension of DIP crystallites along the exciton transport direction, i.e. normal to the (001) plane (fig \ref{fig:fig1}). For thick films ($d > 100nm$) non-radiative as well as radiative recombination processes take place during exciton transport within a single crystalline domain, due to a substantial energy barrier at the grain boundaries. However, for thin films ($d < 100nm$), where most of the crystallites will stretch across the entire film, exciton transport and recombination are substantially determined by the respective quenching interface. Consequently, the variation with temperature is mainly governed by the $PL^{Q}$ contribution to Q. Furthermore this model is supported by the fact that the temperature independent Q below 80 K is significantly smaller for thin than for thick DIP films as quenching is still present in case of the former whereas in thick films excitons are not being capable of reaching the quenching DIP/CuPc interface and therefore relax to the ground state by the same mechanisms as for the uncovered DIP section. 
\\
Finally, this model also explains the non-vanishing quenching contribution for very thick films (fig. \ref{fig:fig2}). In this case, the intrinsic crystallite size, which agrees very well with the optimized thickness deduced from the $I_{spec} / I_{diff}$ ratio measured by X-ray (fig. \ref{fig:fig1}b), initiates a pronounced 3D island growth (Volmer-Weber growth) together with an enhanced surface roughening as confirmed by detailed x-ray analyses and atomic force microscopy (AFM) \cite{duurrschreiber}. As indicated in fig. \ref{fig:fig1}c this topography promotes the penetration of the top-deposited molecular quenching material into the grain boundaries, leading to a significant contribution by lateral exciton diffusion and dissociation to the PL quenching of about 50 \% even for thicknesses larger than 250 nm. In addition, this trend is supported by the better overlap of the electronic wavefunction and the molecular transition dipoles within the (001) plane, presumably entailing the lateral components of the tensorial diffusion constant to be larger than the out-of-plane component \cite{{luntgiebink},{heinemeyerscholz}}. A lower limit for the lateral $L_D$ of several tens of nanometres has been reported which overall matches our observation \cite{zhangmeixner}. However care has to be taken as those values have been determined in the presence of metallic nanostructures which might influence the exciton dynamics e.g. by plasmon mediated non-radiative recombination.
\\
Summarizing these results, we suggest a relation to account for the morphological impact of real quencher interfaces, thereby resembling the aforementioned complementary error function 

\bea
Q' = Q&\Big(&  R+(1-R)\cdot erfc\left(\frac{d-d_0}{\sigma}\right)\Big)\
\eea

The average crystallite height and surface roughness are described by d$_0$ and $\sigma$ whereas R qualitatively contains all effects which limit the relative quenching in case of thick layers. Modelling the thickness dependent Q by our extended exciton diffusion model, comprising effects by optical interference and quencher interface morphology, reveals the fit shown in fig. \ref{fig:fig2} and an $L_D$ of 90 nm for DIP. This value is in good agreement with previous reports using spectral photocurrent measurements interpreted by a model of Feng and Ghosh \cite{{kurrlepflaum},{ghoshfeng}}. Comparing the estimated average crystallite height of $d_0$ = 85 nm with $L_D$ renders the correlation between these properties to be evident. Furthermore, the adjusted reflection coefficient ratio between the free and the CuPc-covered DIP volume can be deduced by fitting the measured data points under consideration of eq.(\ref{eq::I}) and amounts to 1.4. Though influenced by the respective interface morphology, this result as well as the relative phase shift $\delta$ almost matches $\pi$/2 which is theoretically expected for a standing-wave with open end (organic/air interface) versus that with fixed end (organic/quencher interface). Finally, the discussed transport model also renders the thickness dependent interface roughness which according to the rapid roughening model observed for DIP scales as $\sigma$=$d^{\beta}$ with $\beta$=0.75 \cite{duurrschreiber}.
\\
In summary, we have investigated the exciton transport in long-range ordered molecular thin films by PL quenching, explicitly accounting for interference effects as well as structural properties at the respective quencher interface. Based on a diffusion model in combination with complementary morphological analysis we were able to describe the thickness and temperature dependent PL quenching behavior of DIP. The resulting thermal transport behavior indicated coherent exciton transport in a temperature regime below 80 K. The diffusion length of 90 nm is in striking agreement with the average DIP crystallite height and, with prospect on implementation in organic photovoltaics, indicates the decisive role of structural order on the exciton transport in molecular semiconductors.
\\
Financial support by Deutsche Forschungsgemeinschaft (project No. PF385/4) and BASF is acknowleged.

\end{document}